\documentclass[twocolumn]{aastex63}

\usepackage{amssymb}
\usepackage{multirow}
\usepackage{todonotes}
\usepackage{yfonts}
\usepackage{amsmath}
\usepackage{tabularx}
\usepackage[figuresright]{rotating}
\usepackage{wrapfig}
\usepackage{epstopdf}
\usepackage{makecell}
\usepackage{longtable}
\usepackage{enumitem}
\usepackage{comment}
\usepackage{lineno}

\usepackage{xcolor}
\definecolor{tndpurple}{RGB}{153,0,153}

\usepackage{changepage}

\newcolumntype{Y}{>{\centering\arraybackslash}X} 

\received{03 21, 2022}
\revised{08 09, 2022}
\accepted{08 31, 2022}

\submitjournal{ApJ}

\shorttitle{Variable star classification with a Multiple-Input Neural Network}
\shortauthors{Szklen\'ar et al.}

\begin{document}


\title{Variable star classification with a Multiple-Input Neural Network}

\author[0000-0002-5610-7697]{T. Szklen\'ar}
\email{szklenar.tamas@csfk.org}

\affiliation{Konkoly Observatory, Research Centre for Astronomy and Earth Sciences (ELKH),\\ H-1121 Budapest, Konkoly Thege Mikl\'os \'ut 15-17, Hungary\\}
\affiliation{CSFK, MTA Centre of Excellence, H-1121 Budapest, Konkoly Thege Mikl\'os \'ut 15-17, Hungary\\}
\affiliation{MTA CSFK Lend\"ulet Near-Field Cosmology Research Group\\}

\author[0000-0002-8585-4544]{A. B\'odi}
\affiliation{Konkoly Observatory, Research Centre for Astronomy and Earth Sciences (ELKH),\\ H-1121 Budapest, Konkoly Thege Mikl\'os \'ut 15-17, Hungary\\}
\affiliation{CSFK, MTA Centre of Excellence, H-1121 Budapest, Konkoly Thege Mikl\'os \'ut 15-17, Hungary\\}
\affiliation{MTA CSFK Lend\"ulet Near-Field Cosmology Research Group\\}

\author[0000-0003-3759-7616]{D. Tarczay-Neh\'ez}
\affiliation{Konkoly Observatory, Research Centre for Astronomy and Earth Sciences (ELKH),\\ H-1121 Budapest, Konkoly Thege Mikl\'os \'ut 15-17, Hungary\\}
\affiliation{CSFK, MTA Centre of Excellence, H-1121 Budapest, Konkoly Thege Mikl\'os \'ut 15-17, Hungary\\}
\affiliation{MTA CSFK Lend\"ulet Near-Field Cosmology Research Group\\}

\author[0000-0002-6471-8607]{K. Vida}
\affiliation{Konkoly Observatory, Research Centre for Astronomy and Earth Sciences (ELKH),\\ H-1121 Budapest, Konkoly Thege Mikl\'os \'ut 15-17, Hungary\\}
\affiliation{CSFK, MTA Centre of Excellence, H-1121 Budapest, Konkoly Thege Mikl\'os \'ut 15-17, Hungary\\}
\affiliation{MTA CSFK Lend\"ulet Near-Field Cosmology Research Group\\}
\affiliation{ELTE E\"otv\"os Lor\'and University, Institute of Physics, Budapest, Hungary\\}

\author[0000-0002-0686-7479]{Gy. Mez\H o}
\affiliation{Konkoly Observatory, Research Centre for Astronomy and Earth Sciences (ELKH),\\ H-1121 Budapest, Konkoly Thege Mikl\'os \'ut 15-17, Hungary\\}
\affiliation{CSFK, MTA Centre of Excellence, H-1121 Budapest, Konkoly Thege Mikl\'os \'ut 15-17, Hungary\\}

\author[0000-0002-3258-1909]{R. Szab\'o}
\affiliation{Konkoly Observatory, Research Centre for Astronomy and Earth Sciences (ELKH),\\ H-1121 Budapest, Konkoly Thege Mikl\'os \'ut 15-17, Hungary\\}
\affiliation{CSFK, MTA Centre of Excellence, H-1121 Budapest, Konkoly Thege Mikl\'os \'ut 15-17, Hungary\\}
\affiliation{MTA CSFK Lend\"ulet Near-Field Cosmology Research Group\\}
\affiliation{ELTE E\"otv\"os Lor\'and University, Institute of Physics, Budapest, Hungary\\}




\begin{abstract}
In this experiment, we created a Multiple-Input Neural Network, consisting of Convolutional and Multi-layer Neural Networks. With this setup the selected highest-performing neural network was able to distinguish 
variable stars based on the visual characteristics of their light curves, while taking also into account additional numerical information (e.g. period, reddening-free brightness) to differentiate visually similar light curves. The network was trained and tested on OGLE-III data using all OGLE-III observation fields, phase-folded light curves and period data. The neural network yielded accuracies of 89--99\% for most of the main classes (Cepheids, $\delta$ Scutis, eclipsing binaries, RR Lyrae stars, Type-II Cepheids), only the first-overtone Anomalous Cepheids had an accuracy of 45\%. To counteract the large confusion between the first-overtone Anomalous Cepheids and the RRab stars we added the reddening-free brightness as a new input and only stars from the LMC field were retained to have a fixed distance. With this change we improved the neural network's result for the first-overtone Anomalous Cepheids to almost 80\%. Overall, the Multiple-input Neural Network method developed by our team is a promising alternative to existing classification methods.
\end{abstract}


\section{Introduction} 
\label{sec:intro}

Recent sky surveys obtained a vast amount of data that pose previously unseen challenges for astronomers during their analysis. While observations of a few targets can be processed manually, this is not feasible in the case of several or hundreds of thousands of targets. Classification of variable stars is a typical case for utilizing automated data analysis, which can be based on different statistical properties of the light curves like mean, standard deviation, kurtosis, skewness (see e.g. \citealp{Nun2015}), Fourier-decomposition \citep{Kim2016}, color information \citep{Miller2015} or applying different  machine learning methods, including random forest \citep[see e.g.,][]{RF} or deep learning \citep{Zhang21}. 

In our previous paper, we introduced an experiment to classify variable stars based on their light curves as images -- similarly to what a human astronomer would perform \citep[][hereafter referred to as Paper I]{Szklenaretal2020}. For this purpose, we selected five main variable star classes: $\delta$ Scutis, eclipsing binaries, RR Lyraes, and Anomalous- and Type-II Cepheids. This experiment showed that this method is able to classify the different variable types observed by the Optical Gravitational Lensing Experiment (OGLE; \citealt{Udalski2015}) with 77--99\% accuracy for  light curves in the OGLE-III and OGLE-IV databases.

As shown in Paper I, image-based classification of variable stars using a Convolutional Neural Network is a viable method. Although this method can achieve very high accuracy, due the similarity of the phase-folded light curves, to further increase the classification accuracy it requires the usage of additional data. A Recurrent Neural Network (RNN) -- as a standard solution for time-series data -- could help to distinguish similar features of light curves, like as it was done for identifying stellar flares in \citet{VidaK_flares}. The latter uses Kepler light curves, which, unlike the OGLE light curves, are very well sampled, essentially continuous, and do not contain as many large gaps as the light curves we use. Another main difference is that the cadence is uniform for all Kepler targets, while it is inhomogeneous in case of OGLE. Therefore, application of RNN would require extremely meticulous and disproportionately massive pre-processing in our case.
In this paper, we extend our experiment by supplying different physical parameters (e.g. period, magnitude) as an auxiliary input to the classifier networks in the hope of improving their performance. The main goal of this work is to investigate the effectiveness of a Multiple-input Neural Network (MINN) in general, where numerical data is expected to help distinguish the main variable stars, and their sub-types as well.
To achieve this, we used numerical data of periods and magnitudes attached to images of phase-folded light curves of OGLE-III periodic variable stars \citep{Udalski2008}.

This paper is structured as follows. In Section\,\ref{sec:data}, we present our data and the process of data augmentation. Section\,\ref{sec:CNN} describes the MINN, while our results are discussed in Section\,\ref{sec:results_and_discussion}. The paper closes with Section\,\ref{sec:conclusions}, in which we give the concluding remarks on our results, respectively.

\section{Data and methods}
\label{sec:data}

OGLE provides one of the most extensive data set of variable stars with reliable, human expert verified classifications, which is crucial to train and test a machine learning algorithm. In Paper I, a Convolutional Neural Network was constructed to classify the periodic variable stars based on the observations of OGLE. Paper I considered only the light curves obtained in the field of the Large Magellanic Cloud, in order to keep the sample as homogeneous as possible. Similar to Paper I, we primarily used the OGLE-III data set, but now we aimed to extend our data sample with the Small Magellanic Cloud, the Galactic bulge, and the Galactic disk data sets.

\subsection{Observational data}

\begin{table*}
\caption{The number of variable stars (both main and sub-types) collected from the original OGLE-III LMC, SMC, Galactic bulge, and Galactic disk data set. Sub-types with only a few members or mostly noisy light curves were excluded.}
\label{Tab:usedlcs}
\begin{tabular}{ccrrrr|r}
\hline
\multicolumn{1}{c}{} Main type & sub-type & LMC & SMC & Galactic bulge & Galactic disk & Total\\
\hline
\hline
\bf{ACep}   & -         & \bf{83}	    & \bf{\nodata}	& \bf{\nodata} 	& \bf{\nodata} 	& \bf{83}\\
\hline
            & F	        & 62	        & \nodata 	    & \nodata       & \nodata       & 62\\
            & 1O	    & 21	        & \nodata 	    & \nodata	    & \nodata       & 21\\
\hline
\bf{Cep}	& -         & \bf{3\,262}	& \bf{4\,485}	& \bf{28}	    & \bf{\nodata} 	& \bf{7\,775}\\
\hline
            & F	        & 1\,818	    & 2\,626	    & 21	        & \nodata       & 4\,465\\
            & 1O	    & 1\,238	    & 1\,644	    & 4	            & \nodata       & 2\,886\\
            & 1O2O	    & 206	        & 215	        & 3 	        & \nodata       & 424\\
\hline
\bf{DSct}   & -         & \bf{2\,788}	& \bf{\nodata} 	& \bf{\nodata} 	& \bf{\nodata} 	& \bf{2\,788}\\
\hline
            & SINGLEMODE& 2\,696	    & \nodata       & \nodata       & \nodata       & 2\,696\\
            & MULTIMODE	& 92	        & \nodata       & \nodata       & \nodata	    & 92\\
\hline
\bf{ECL}	& -         & \bf{23\,993}	& \bf{6\,138}	& \bf{\nodata}	& \bf{7\,434}	& \bf{37\,565}\\
\hline
  	        & EC	    & 1\,048	    & 777	        & \nodata       & 7\,434	    & 9\,259\\
 	        & ED	    & 16\,443	    & 5\,361	    & \nodata       & \nodata       & 21\,804\\
 	        & ESD	    & 6\,502	    & \nodata       & \nodata       & \nodata       & 6\,502\\
\hline
\bf{RRLyr}	& -	        & \bf{23\,637}	& \bf{2\,366}	& \bf{16\,835}	& \bf{\nodata}	& \bf{42\,838}\\
\hline
 	        & RRAB	    & 17\,693	    & 1\,933	    & 11\,755	    & \nodata       & 31\,381\\
 	        & RRC	    & 4\,958	    & 175	        & 4\,989	    & \nodata       & 10\,122\\
 	        & RRD	    & 986	        & 258	        & 91	        & \nodata       & 1\,335\\
\hline
\bf{T2Cep}	& -	        & \bf{186}	    & \bf{36}	    & \bf{357}	    & \bf{\nodata}	& \bf{579}\\
\hline
 	        & BLHer	    & 64	        & 17	        & 156	        & \nodata       & 237\\
 	        & RVTau	    & 42	        & 9	            & 73	        & \nodata       & 124\\
 	        & WVir	    & 80	        & 10	        & 128	        & \nodata       & 218\\
\hline
\hline
\end{tabular}
\end{table*}

\begin{figure}
   \centering
   \includegraphics[width=\columnwidth]{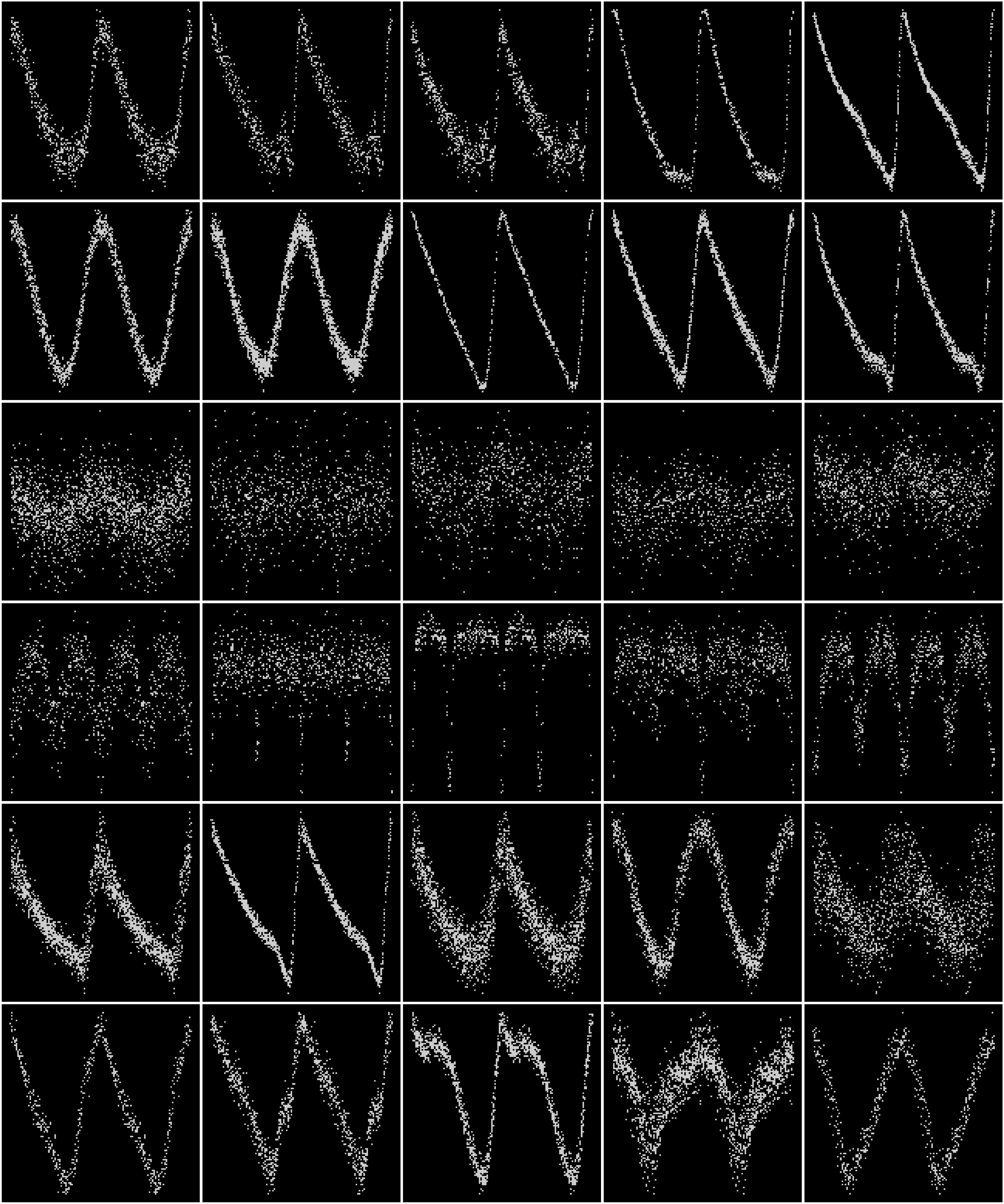}
      \caption{Gallery of light curve images of the different variable stars types used in this work. The light curves are from the OGLE-III catalog; pulsating variable stars and eclipsing binaries are phase folded in the [0..2] interval by their pulsating periods and the orbital periods, respectively. Images, from top to bottom, in each row represent ACep, Cep, DSct, ECL, RRLyr and T2Cep variable stars. This gallery only shows light curves with real measurements and they were resized and have 128x128 pixel resolution as it was used in the first input of the neural network.} 
      \label{fig:lcurves}
\end{figure}

The OGLE-III catalog lists more than 100\,000 variable stars, containing measurements from the Galactic bulge, the Galactic disk, and the Magellanic Clouds. The observations were obtained in the $I$ and $V$ bands. As the amount of $I$ band observations exceeds many times  the $V$ band measurements (about 15 times more), we chose to work with the $I$ band data only.
Along with the photometric observations, the OGLE-III catalog contains some fundamental parameters of the objects (e.g., periods, amplitudes, colors). We collected these values for every variable star presented in our research. From the available information, we utilized the periods and calculated the Wesenheit-index \citep[][]{StarsAndStellarSystems}:
\begin{equation}
    W = I - 1.55 (V-I),
\end{equation}
which served us as an additional input parameter for the classification process.

In this work, we focus on 6 different main variable star types: Anomalous Cepheids \citep[ACep,][]{OGLEIII_ACep_T2Cep}, Classical Cepheids \citep[Cep,][]{OGLEIII_CEP_LMC, OGLEIII_CEP_SMC, OGLEIII_CEP_T2CEP_BLG}, $\delta$ Scutis \citep[DSct,][]{OGLEIII_DSCT}, eclipsing binaries \citep[ECL,][]{OGLEIII_ECL_LMC, OGLEIII_ECL_SMC, OGLEIII_ECL_BLG, OGLEIII_ECL_GD}, RR Lyrae stars \citep[RRLyr,][]{OGLEIII_RRLYR_LMC, OGLEIII_RRLYR_SMC, OGLEIII_RRLYR_BLG}, and Type-II Cepheids \citep[T2Cep,][]{OGLEIII_ACep_T2Cep, OGLEIII_T2CEP_SMC, OGLEIII_CEP_T2CEP_BLG}.
The main variable star classes are divided into several sub-classes, excluding those which have only a few members or contain mostly noisy light curves.
Table\,\ref{Tab:usedlcs} lists the final number of  variable stars per classes collected from the OGLE-III database. We note in passing that we used Classical Cepheids (Cep) in this work, while we omitted them in Paper~I.

Using the epochs and periods from the OGLE-III catalog, the light curves have been phase-folded and transformed into 1 bit (black and white) images with a size of $512\times512$ pixels (see Figure\,\ref{fig:lcurves}). To phase-fold the light curves of pulsating variables we used the pulsation periods, while for eclipsing binaries we used the orbital periods (i.e., twice of the formal periods). 
The horizontal range of the phase-folded light curve is set by the (main) period of the periodic variable star (differs star by star), such that all plots will have a range of [0,1], while the y scale is set by the amplitude (range of light variation). 
To clean the data set, phase-folded light curves were fitted with a Savitzky--Golay filter \citep[][]{Savitzky_Golay}. Afterward, points that were further than three standard deviations away from the mean of the residual light curves have been excluded. 

\subsection{Constructing the training and testing sample}
\label{sec:data_aug}

\begin{table*}
\footnotesize
\caption{ The number of original and generated variable star light curves used to train and test our neural network in case of the two setups (using only the main types and separating stars into different sub-types), first using all OGLE-III fields and after that only the LMC field to be able to utilize brightness for stars with known distances.
}
\label{Tab:lc_generation}
\begin{center}
\begin{tabular}{ccccc|cc}
\hline
\multicolumn{5}{c|}{Light curves from all OGLE-III fields} & \multicolumn{2}{c}{OGLE-III LMC field only} \\
\hline
Main type & Sub-type & Original$^\dagger$ & Main type sample & Sub-type sample & LMC originals & LMC sub-type sample \\
& & & \textit{Training/Testing} & \textit{Training/Testing} & & \textit{Training/Testing}\\
\hline
\hline
\bf{ACep}   & --      	& \bf{83}       & 9\,000/1\,000     &               & \bf{83}       &  \\
\hline
            & 1O	    & 21            & 2\,277/500        & 4\,500/500    & 21    & 4\,500/500\\
            & F	        & 62            & 6\,723/500        & 4\,500/500    & 62    & 4\,500/500 \\
\hline
\bf{Cep}	& --         & \bf{7\,775}   & 9\,000/1\,500    &               & \bf{3\,262}     & \\
\hline
            & 1O	    & 2\,886        & 3\,341/500        & 4\,500/500    & 1\,818    & 4\,500/500 \\
            & 1O2O	    & 424           & 491/500           & 4\,500/500    & 1\,238    & 4\,500/500 \\
            & F	        & 4\,465        & 5\,168/500        & 4\,500/500    & 206   & 4\,500/500 \\
\hline
\bf{DSct}   & --         & \bf{2\,788}   & 9\,000/1\,000    &               & \bf{2\,788} \\
\hline
            & MULTIMODE& 92            & 297/500           & 4\,500/500    & 2\,696    & 4\,500/500 \\
            & SINGLEMODE& 2\,696       & 8\,703/500        & 4\,500/500    & 92    & 4\,500/500 \\
\hline
\bf{ECL}	& --         & \bf{37\,565}  & 9\,000/1\,500    &               & \bf{23\,993} \\
\hline
            & EC	    & 9\,259        & 2\,218/500        & 4\,500/500    & 1\,048    & 4\,500/500 \\
            & ED	    & 21\,804       & 5\,224/500        & 4\,500/500    & 16\,443   & 4\,500/500 \\
            & ESD	    & 6\,502        & 1\,558/500        & 4\,500/500    & 6\,502    & 4\,500/500 \\
\hline
\bf{RRLyr}	& --	        & \bf{42\,838}  & 9\,000/1\,500     &           & \bf{23\,637} \\
\hline
            & RRAB	    & 31\,381       & 6\,593/500        & 4\,500/500    & 17\,693   & 4\,500/500 \\
            & RRC	    & 10\,122       & 2\,127/500        & 4\,500/500    & 4\,958    & 4\,500/500 \\
            & RRD	    & 1\,335        & 280/500           & 4\,500/500    & 986   & 4\,500/500 \\
\hline
\bf{T2Cep}	& --	        & \bf{579}      & 9\,000/1\,500     &           & \bf{186} \\
\hline
            & BLHer	    & 237           & 3\,684/500        & 4\,500/500    & 64    & 4\,500/500 \\
            & RVTau	    & 124           & 1\,927/500        & 4\,500/500    & 42    & 4\,500/500 \\
            & WVir	    & 218           & 3\,389/500        & 4\,500/500    & 80    & 4\,500/500 \\
\hline
& & \bf{Total used}      & \bf{54\,000/8\,000}  & \bf{72\,000/8\,000} & & \bf{72\,000/8\,000} \\
\hline
\multicolumn{5}{l}{$^\dagger$ The original OGLE-III data set from the LMC, SMC, galactic bulge and disk fields.}
\end{tabular}
\end{center}
\end{table*}

Paper I, as well as others, pointed out the necessity of data augmentation in case of a highly unbalanced data set. In this work, we train and test two kind of setups for which we apply a data augmentation approach, which is different from that used in Paper I. To increase the sample of underrepresented classes, we applied Gaussian Process regression (see Subsection~\ref{sec:GP}).

First, we focused only on the classification of the main variable star classes. In this case, data augmentation was applied to construct a data set consisting of 10\,000 and 10\,500 light curves for classes with two and three sub-types, respectively. During the generation of artificial light curves, the ratio of the various sub-types within a class was taken into account. For example, the original data set contains 83 ACeps, of which 62 are fundamental mode pulsators (F sub-type), while 21 are first overtone pulsators (1O sub-type). In this case, we created an augmented data set in which the ratio of F and 1O sub-types are 7\,470 to 2\,530, representing the original proportion. Table~\ref{Tab:lc_generation} lists the number of original and augmented data sets for each variable star type. Note that in those cases where the number of samples within a given sub-type in the original data set is more than that is required in the final sample (e.g., in case of eclipsing binaries, RR Lyrae stars), data augmentation was not applied. Instead, the desired number of stars were randomly selected from the original data set.

The data set used for training and testing the neural network contained 62\,000 images. 9\,000 light curves were used from each main variable star type for training/validation with a ratio of 70/30\% and 500 light curve images were selected from each variable star sub-type for testing purposes.

Secondly, as much of the sub-types in a given main variable star class show explicitly distinguishable light curve shapes we decided to perform a training using the sub-types separately.

We divided the main classes into sub-types as they are labeled in the OGLE catalog (e.g., Cepheids pulsating in the fundamental mode -- F, first radial overtone -- 1O, etc.). In this case, each sub-type was balanced to contain 5\,000 light curves, either augmenting the data set or randomly selecting light curves from the original sample.
For the training and validation of classification of 16 different labels we sampled 4\,500 images. For testing, 500 light curve images were selected from every variable star sub-types without any overlapping with the training and validation samples. Altogether we used 80\,000 light curves in this article which was enough to train and test both the main types and the sub-types.
For more details, see Table\,\ref{Tab:lc_generation}.

Here we note that, in order to avoid false predictions we ensured that the teaching and testing samples do not overlap. For a given star the original and synthetic light curves are only present in one of the three steps, teaching, validation, or testing.

\subsection{Generating synthetic light curves}
\label{sec:GP}

\begin{figure*}
   \centering
   \includegraphics[width=\textwidth]{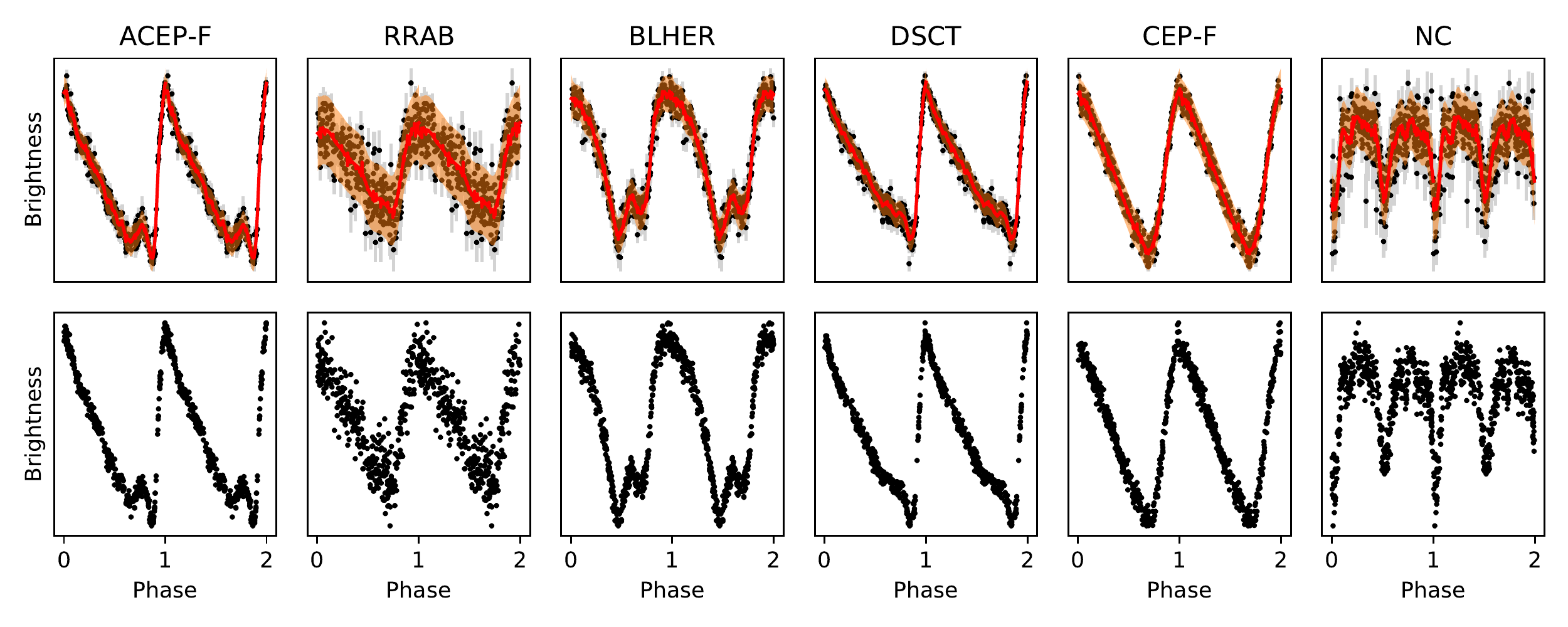}
      \caption{Gallery of some variable star sub-types to show the process of synthetic data generation using Gaussian Process (GP) regression. Top: black points are the original phase-folded light curves with their uncertainties (gray bars). The red lines are the mean of GP predictions, and the shaded orange areas show the three standard deviation from the mean. Bottom: the artificially generated phase-folded light curves that were randomly sampled from the GP posteriors at random phase values.}
      \label{fig:generatedLCs}
\end{figure*}

\begin{figure}
   \centering
   \includegraphics[width=\columnwidth]{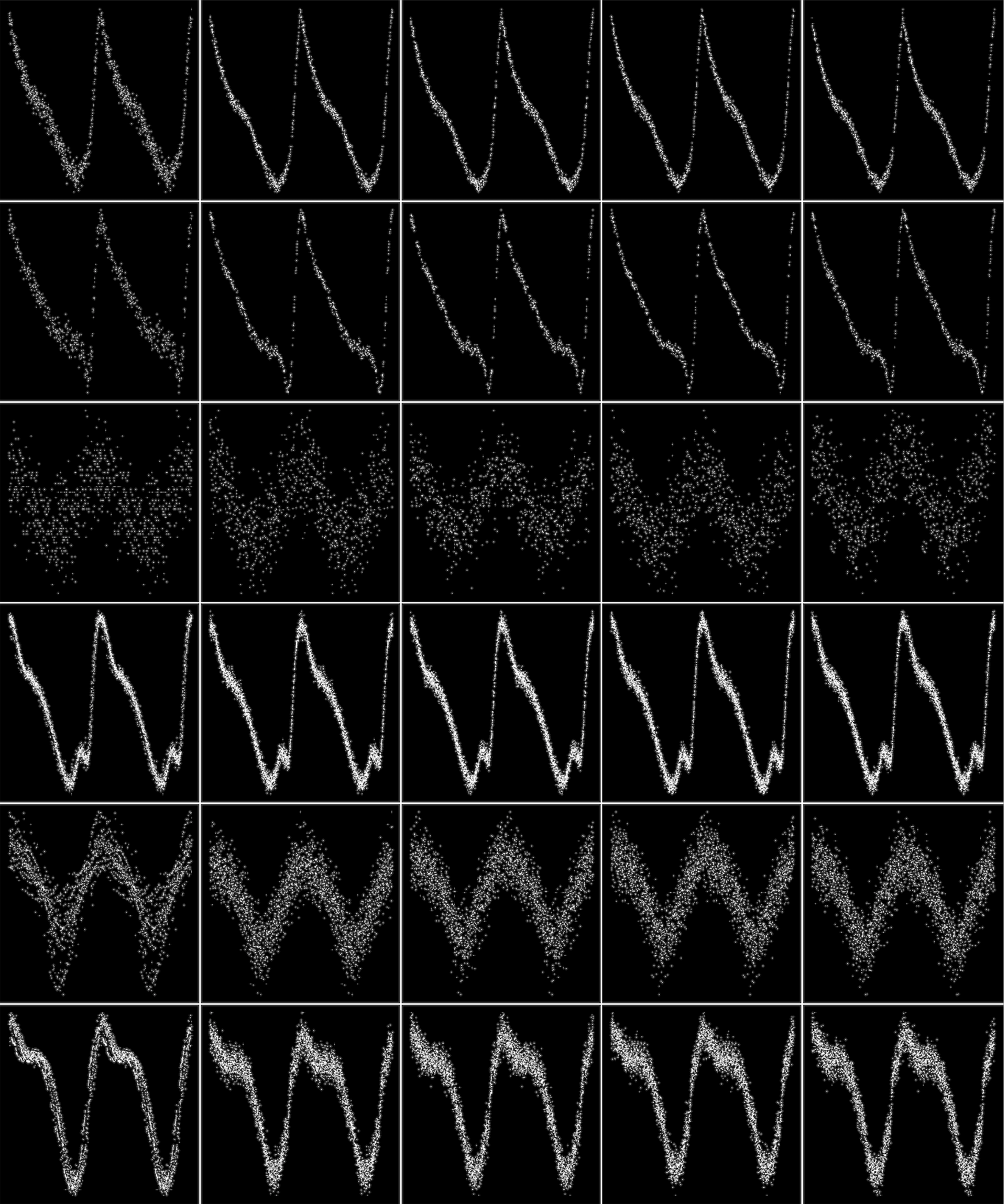}
      \caption{Gallery of artificially generated phase-folded light curve images. The first column shows the original light curves, of which the synthetic data were generated. The rows correspond to the following variable star sub-types: ACep~1O, ACep~F, DSct Multimode, T2Cep BLHer, T2Cep, RVTau and T2Cep WVir. Most of the artificial light curves were generated for these sub-types. The light curve images shown in this figure were resized and have 128x128 pixel resolution as they were used in the neural network.} 
      \label{fig:dummyLC}
\end{figure}

To generate synthetic light curves, a physical model is needed for each variable type represented in our sample. As we lack such a model set, we need a method that is flexible enough to model the different light curve shapes, and provide reliable uncertainties. To overcome this problem, the Gaussian Process (GP) regression is used.

GPs are stochastic, continuous, non-parametric models \citep{GPbook}. GP is a distribution over functions, which is fully described by its mean and a covariance matrix or kernel function. The data is represented with a general multivariate Gaussian distribution
\begin{equation}
    p(\mathbf{m}|\mathbf{t}, \mathbf{\alpha}) = \mathcal{N}(\mathbf{\mu}(\mathbf{t}), \mathbf{K}(\mathbf{t},\mathbf{\alpha}))
\end{equation}
where $\mathbf{m}(\mathbf{t})$ is the time series of observations, $\mathbf{m}$ and $\mathbf{t}$ are the vectors of fluxes and time, respectively. $\mathcal{N}$ depicts a
Gaussian distribution with the mean function $\mu(\mathrm{t})$ and covariance matrix $\mathrm{\mathbf{K}}(\mathrm{\mathbf{t}}, \mathbf{\alpha})$, and $\mathbf{\alpha}$ is a vector of the hyperparameters characterizing the covariance matrix.
The mean function can be set to any function, however, it is often considered to be 0. The kernel function describes the covariance
between any pair of the points drawn from a given GP. Usually a kernel function assumes that the covariance between points
is a function of the distance between the points of the independent variable, which is the time (or phase) in case of a light curve.

The covariance matrix, $\mathrm{\mathbf{K}}$, is defined as:
\begin{equation}
    K_{ij} = \sigma^2_i \delta_{ij} + k_\alpha(\tau_{ij}),
\end{equation}
where $\sigma^2_i$ is the measurement error given for the $i$\textsuperscript{th} observation, $\delta_{ij}$ is the Kronecker delta function and $k_\alpha(\tau_{ij})$ is the kernel with $\tau_{ij}= |t_i - t_j|$, the distance between $i$\textsuperscript{th} and $j$\textsuperscript{th} time points.

The GP regression optimizes the set of parameters $\alpha$ by minimizing the log-likelihood function
\begin{equation}
\ln \mathcal{L}(\mathbf{\alpha}) = -\frac{1}{2}\mathbf{r}^{T} \mathbf{K}^{-1} \mathbf{r} -\frac{1}{2} \ln |\mathbf{K}| -\frac{N}{2}\ln(2\pi) ,
\end{equation}
where $\mathbf{r}$ is the residual after subtracting the mean model from the observations, and N is the number of data points.

As the computational cost of GP regression scales with $\mathcal{O}(n^3)$, we have to choose an implementation, which is computationally efficient. In this work we used the GP implemented in the \texttt{exoplanet}\footnote{\texttt{exoplanet} is a toolkit for probabilistic modeling of astronomical time series, which is built upon \texttt{theano} \citep{exoplanet:theano}, \texttt{PyMC3} \citep{exoplanet:pymc3} and \texttt{celerite} \citep{celerite}.} python package \citep{exoplanet}.

The first step of GP modeling is to choose an adequate kernel. In the \texttt{exoplanet} implementation the kernel is a mixture of exponential functions:
\begin{equation} \label{kalpha}
    k_\alpha(\tau_{ij}) = \sum_{m=1}^M a_m \exp(-c_m \tau_{ij}).
\end{equation}
If the $a_m$ and $c_m$ parameters are complex numbers, $a_m \rightarrow a_m \pm i b_m$, $c_m \rightarrow c_m \pm i d_m$, then Equation~\ref{kalpha} can be rewritten as a sum of sine of cosine terms and the result is a mixture of quasi-periodic oscillators:
\begin{equation} \label{eq:kalphatau}
\begin{split}
k_\alpha(\tau_{ij}) = \sum_{m=1}^M &[a_m \exp(-c_m \tau_{ij}) \cos(d_m \tau_{ij}) \\
&+ b_m \exp(-c_m \tau_{ij}) \sin(d_m \tau_{ij}) ],
\end{split}
\end{equation}
where the parameter set $\alpha = \{a_m, b_m, c_m, d_m\}$.

From the available kernels we chose the RotationTerm, which is a mixture of two SHO terms, which can be used to model stochastic variability in a time series. In the Fourier space, the SHO term representing a stochastically-driven, damped harmonic oscillator with power spectral density \citep{celeriteGP}:
\begin{equation} \label{eq:Somega}
    S(\omega) = \sqrt{\frac{2}{\pi}} \frac{S_0 \omega_0^4}{ (\omega^2-\omega_0^2)^2 + \omega_0^2\omega^2/Q^2 }, 
\end{equation}
where $\omega$ is an angular frequency, $\omega_0$ is the frequency of the
undamped oscillator, $Q$ is the oscillator’s quality factor, and
$S_0$ is proportional to the power at $\omega = \omega_0$,
\begin{equation}
S(\omega_0) = \sqrt{\frac{2}{\pi}} S_0 Q^2.
\end{equation}
\citet{celeriteGP} showed that if the parameters are chosen properly, then Equation~\ref{eq:Somega} can be match to Equation~\ref{eq:kalphatau}, of which the kernel can be rewritten as
\begin{equation}
\begin{split}
k_\mathrm{SHO}(\tau;S_0,Q,\omega_0) = S_0 \omega_0 Q \exp{(-\frac{\omega_0\tau}{2Q})} \times \\
\begin{cases}
\cosh(\eta \omega_0 \tau) + \frac{1}{2\eta Q} \sinh(\eta \omega_0 \tau), & 0 < Q < 1/2 \\
2(1 + \omega_0 \tau), & Q = 1/2 ,\\
\cos(\eta \omega_0 \tau) + \frac{1}{2\eta Q} \sin(\eta \omega_0 \tau), & 1/2 < Q
\end{cases}
\end{split}
\end{equation}
where $\eta = |1-(4Q^2)^{-1}|^{1/2}$.

The goal of the GP regression is to represent the different light curve shapes with models which confidence intervals can be used to sample new data sets from the original measurements. As, instead of the time series itself, we use the well-sampled phase-folded light curves for the classification, we used the latter for GP regression too. The optimization of the kernel parameters were based on Bayesian parameter estimation with normally distributed priors. As we only want to represent the different light curve shapes and are not interested in the actual parameter values and its uncertainties, to minimize the log-likelihood, we used the \texttt{minimize} method from \texttt{scipy.optimize} \citep{2020SciPy-NMeth}.

For each underrepresented variable star, where we needed to augment the data set, we randomly sampled new points from the GP posteriors at random phase values. The number of new points were the same as the original ones. Some example phase-folded light curves, along with their GP fits and the synthetic data sets can be seen in Figure \ref{fig:generatedLCs}. Figure \ref{fig:dummyLC} shows a gallery of artificially generated light curve images for sub-types, where data augmentation was needed.

\subsection{Numerical data}
\label{sec:used_data_set_num}

\begin{figure}
   \centering
   \includegraphics[width=\columnwidth]{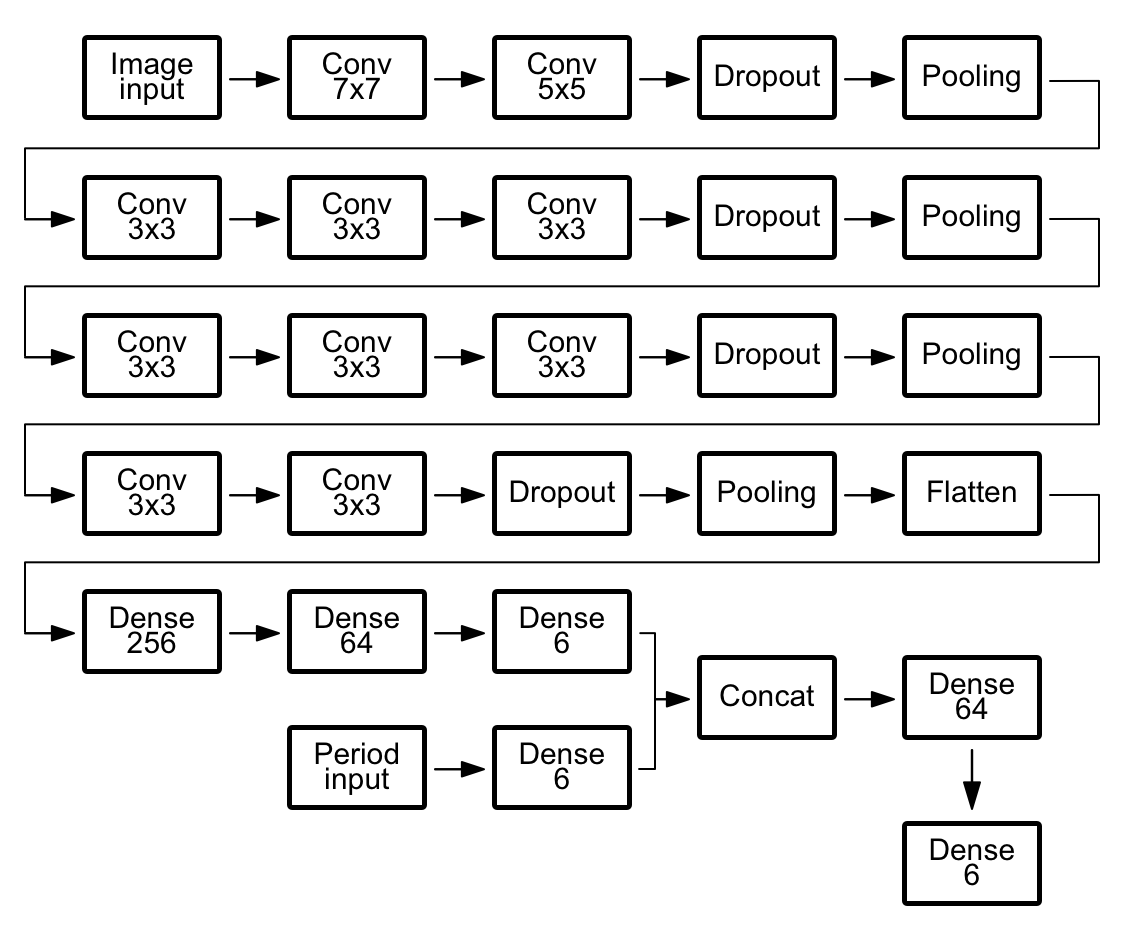}
      \caption{Schematic view of the architecture of the Multiple-input Neural Network built in this study. In this case two inputs are present, the first one is the data set of phase-folded light curve images, the second one is the numerical data of the periods associated with the given stars. This figure shows the neural network used to classify stars into main variable star classes. When we classified stars into sub-classes, the number of the neurons in the last dense layer and those dense layers that are right before the concat layer are changed to 16.}
    \label{fig:arch_imgP}
\end{figure}

\begin{figure}
   \centering
   \includegraphics[width=\columnwidth]{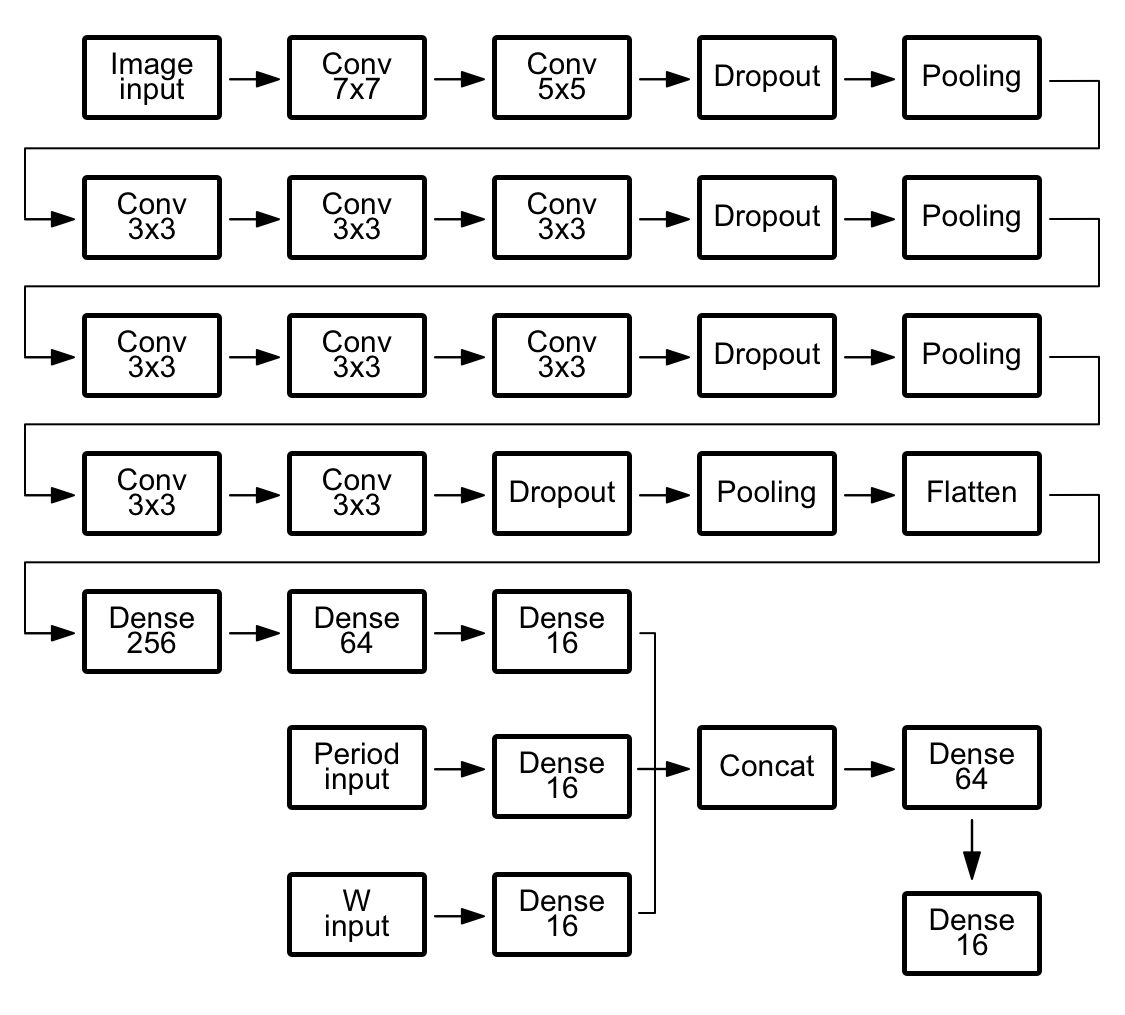}
      \caption{The architecture of our extended Multiple-input Neural Network, where the inputs are the following: phase-folded images, the period and Wesenheit-index of the variable stars. We used this architecture to classify stars observed in the LMC into 16 different variable star sub-classes.}
         \label{fig:arch_imgPW}
   \end{figure}

Due to the similarities of the light curve shapes, the classification using only the light curve images led to false predictions. We needed additional data to be able to distinguish the different variable stars with even higher accuracy. We created a data file containing every numerical data of the objects used in our project downloaded from the OGLE-III database\footnote{\url{https://ogledb.astrouw.edu.pl/~ogle/CVS/}}. These numerical parameters were the star's pulsation or orbital period, apparent magnitude, and amplitude. This sample was extended with the Wesenheit-index.

We worked with 80\,000 images and the same amount of numerical data in the final data set. For the augmented light curves, the same physical parameters were given as for the original data set.

\section{Multi-Input Neural Network}
\label{sec:CNN}

This work relies on Paper I which gives us a solid base to try to improve the performance of the Convolutional Neural Network (CNN) and integrate it with new features. The neural network in Paper I used phase-folded light curve images as inputs. Although this CNN worked reliably, the similarity of light curves belonging to different variable star classes led to false predictions. As in this paper the light curves are augmented with a different method  we had to make changes to the CNN to be able to extract the fine features. After testing the performance of the CNN we aimed to create a neural network in which - besides the CNN - additional numerical inputs can be used for the classification process. We called this latter architecture as Multiple-Input Neural Network.

\begin{figure*}
  \centering
  \includegraphics[width=\textwidth]{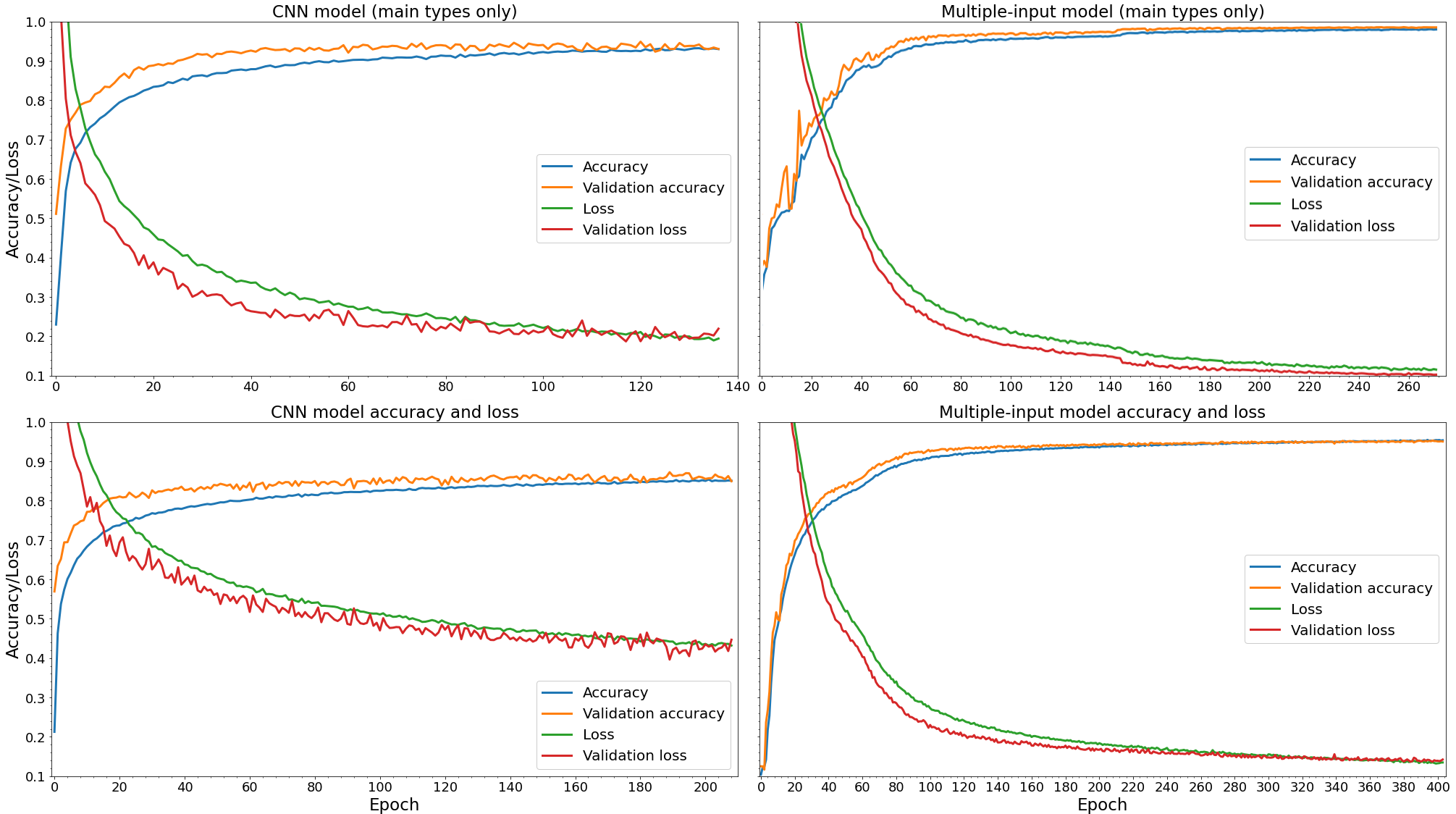}
      \caption{ Performance comparison of the CNN (left column) and Multiple-Input Neural Network (right column) showing the neural networks' accuracy and loss evolution during the training and validation phases. The first row shows the training and validation performance with the 6 main types, while the bottom row shows the neural network's performance with $16$ variable star sub-types. Due to the additional numerical data, the Multiple-input Neural Network outperforms the CNN. At the end of the training phase it reaches much better accuracy and performs more reliable variable star sub-type classification. Note that the x-scale is different in the sub-panels.}
         \label{fig:models_acc_loss_history}
   \end{figure*}

 \begin{figure}
   \centering
   \includegraphics[width=\columnwidth]{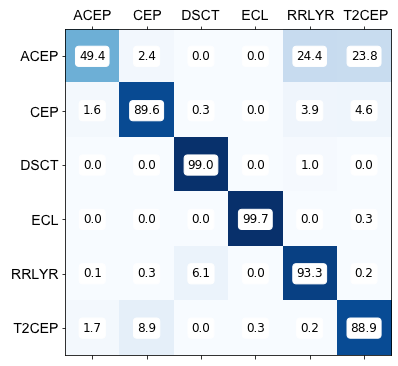}
      \caption{Confusion matrix of the classification result of the test sample of six main variable star classes. Light curve images and the periods were used as inputs for the multiple-input neural network.}
         \label{fig:multi_main}
   \end{figure}

\subsection{Architecture of the neural network}
\label{sec:neural_network_architecture}

The network was developed using the Keras API built over TensorFlow \citep{Tensorflow}, an open source platform for machine learning. For detailed description of the different layers used in the network, see Paper I.

First we designed a multi-input neural network to classify the different variable stars observed in all OGLE-III fields into 6 main and 16 sub-classes, separately. The architecture of this network, which uses two different inputs (images and numerical data), can be seen in Figure\,\ref{fig:arch_imgP}.

After some experimentation we retained only the period as the most informative parameter as additional input in this neural network.

In the case of the Large Magellanic Cloud (LMC), the neural network was extended to handle the Wesenheit-index as an additional input in order to classify variable star with similar distances (i.e., absolute magnitudes) into sub-types.
This architecture can be seen in Figure\,\ref{fig:arch_imgPW}.

\subsection{Image classification}
\label{sec:CNN_creation}

The architecture of the image classification section of the network can be separated into 4 different parts. The first part consists of two convolutional layers with $7\times$7 and $5\times$5 convolutional windows, followed by a dropout and a pooling layer. The purposes of this first section is the extraction of low-level features and the resizing of input images.

The second part contains two blocks which are exactly the same, having three convolutional layers using $3\times$3 convolutional windows, ending with one dropout and one pooling layer. In these layers the high-level features are extracted.

The third section has only two convolutional layers with the usual dropout and pooling layers. The output matrix reaches its minimum size after the last pooling layer, where we apply a flatten layer to create a usable input for the last section.

In the fourth section we use only fully-connected (FC) layers with decreasing number of units. The last FC layer applies a softmax activation to classify the images into separate variable stars classes or sub-classes.

\begin{figure}
   \centering
   \includegraphics[width=0.8\columnwidth]{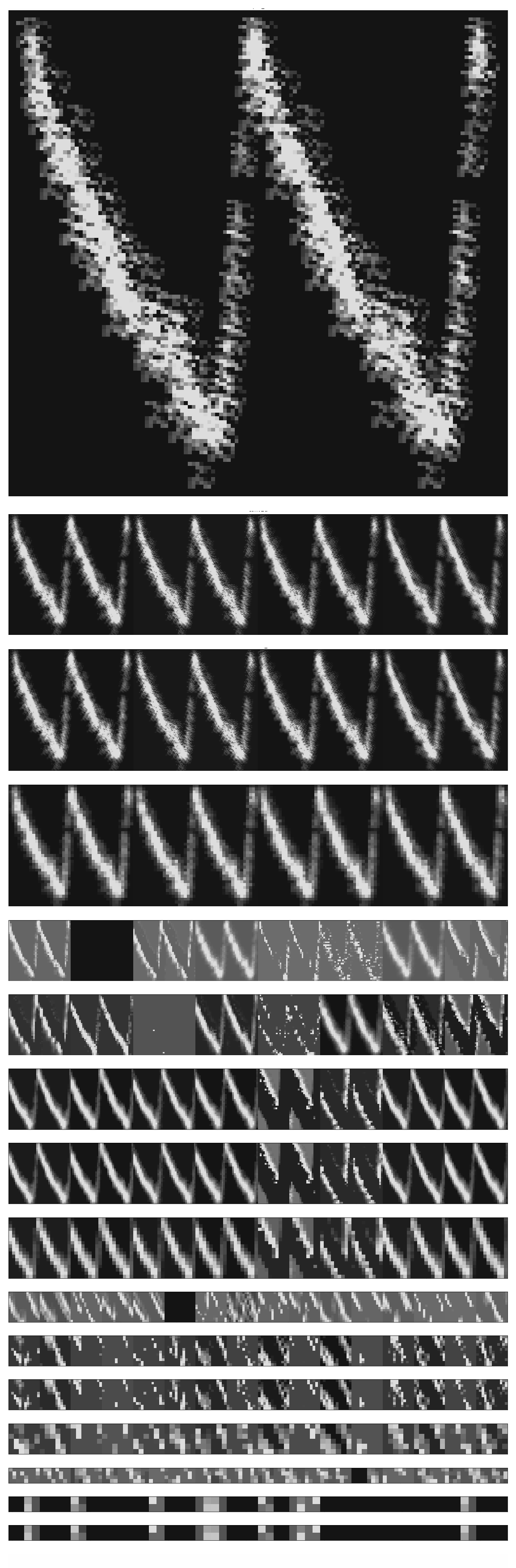}
      \caption{ Features learned by the convolution neural network (part of the MINN), visualized by images of the latent spaces of the convolutional layers.
}
         \label{fig:cnn_visualization}
   \end{figure}

The original resolution of the generated light curve images was $512\times512$ pixels, which was reduced to $128\times128$ pixels during the read-in of the files and the pixel values were converted to integer numbers. These greatly improved the running time of the algorithm while  preserving the information content and yielding the same performance.
Figure\,\ref{fig:cnn_visualization} shows features learned by the convolution neural network (part of the MINN), visualized by images of the latent spaces of the convolutional layers.

\subsection{Handling numerical data}
\label{sec:numerical_data}

As discussed in Section~\ref{sec:used_data_set_num}, in order to improve the effectiveness of the classification process, we aimed to include additional data for each variable star. OGLE-III database contains basic physical parameters for each target. These are basically floating point numbers, which can be used as input parameters without any preprocessing in the neural network.
Thus, a simple dense layer with softmax activation was applied after the read-in of the numerical input, which classifies the stars based on their periods/brightness into separate variable star classes or sub-classes. The size of the output changed according to the number of examined types.

\subsection{Concatenating the outputs}
\label{sec:concat}
As both the image and the numerical data have the same number of output classes, we can concatenate them and use the result as a new input for further classification. The two (or more) concatenated outputs will be handled as a single input into a fully-connected layer with 64 units and this will be sent to a softmax dense layer which will make the final classification.

\begin{table}
\footnotesize
\centering
\caption{Hyperparameters of our Multiple-input Neural Network} 
\label{Tab:hyperparameters}
\begin{tabular}{rcc}
\hline
Parameter & Tested values & Chosen value \\
\hline
\hline

\noalign{\smallskip}
\multicolumn{3}{c}{\emph{Architecture}} \\
\noalign{\smallskip}

Starting convolution window & [$3\times3]$&\\
                            & [$5\times5]$ & [$7\times7]$\\
                            & [$7\times7]$ &\\
Convolution stride & [1,2,3] & 1\\
Convolution padding & 0 & 0\\
Convolution activation &  ELU \\& ReLU \\& Leaky ReLU & ReLU \\
Dropout probability & [0.1--0.5] & 0.3 \\
Pooling type & MaxPooling & MaxPooling \\
Pooling size & [$2\times2$, $3\times3$] & $2\times2$ \\
Number of convolution layers & 10 & 10 \\
Number of pooling layers & 4 & 4\\
Number of dense layers & 6 & 6\\
Dense activation function & ReLU & ReLU\\ 

\hline\noalign{\smallskip}
\multicolumn{3}{c}{\emph{Optimization}} \\
\noalign{\smallskip}

Batch size & [32 - 2048] & 256 \\
Learning rate &  [$10^{-1}$--$10^{-5}$] & $2.5\times10^{-4}$\\
Optimizer & \multicolumn{2}{c}{Adam}  \\
Loss function & \multicolumn{2}{c}{Categorical crossentropy}\\
Early stopping $\Delta$ and patience &
\multicolumn{2}{c}{$10^{-4}$, 19 epochs}  \\
\hline
\end{tabular}
\end{table}

\subsection{Optimizer, learning rate and batch size} 
\label{sec:optimizers}
We tested in detail the performance of the Adam (Adaptive Moment estimation) optimizer with various setups, changing the learning rate between $10^{-1}$ and $10^{-5}$ and the batch size in the range of 32 to 2048. We found that our model performed the best with a learning rate of $0.00025$ and a batch value of 256, thus we chose this parameter set in our final model. Table\,\ref{Tab:hyperparameters} lists the tested and best hyperparameters of the neural network.

\subsection{Early stopping}
\label{sec:earlystopping}
To avoid overfitting, we applied an \texttt{EarlyStopping} callback during training, which monitors the change of the validation loss value. We set the \texttt{min\_delta} and \texttt{patience} values to $10^{-4}$ and 19, respectively. These parameters control the minimum change in the monitored quantity to qualify as an improvement during training, and the number of epochs\footnote{When using the term epoch, we so not refer to the timestamp of the observations but to the well-known machine learning term meaning one cycle during which the algorithm completes one full pass on the training dataset.} with no improvement after which training will be stopped, respectively. With these parameters, the callback monitors the validation loss change and if it does not decrease by at least $10^{-4}$, the callback will run for another 19 additional epochs, stops the training process and saves the best weights for further testing.

\subsection{Performance} 
\label{sec:performance}
For the training and testing process, we used a GPU-accelerated computer containing NVidia GeForce RTX 2080 Ti GPU cards. The training phase usually took about 300 epochs, where one epoch lasted for 6 seconds. The whole training and validation phase took about 1.5 hours. The classification of the complete test data set (8000 variable stars) took approximately 3 seconds.

\subsection{K-Fold cross-validation}
\label{sec:kfold}

As the performance of a neural network is highly dependent on the training set, we decided to carry out a K-fold cross-validation test to quantify the reliability of our CNN classification results. Here, the first step is to separate a subset from the whole data set, which is used for the testing. The remaining data set is separated into $k$ distinct parts with equal sizes. From these a single subset is used for testing, and the last $k-1$ sets are used for training the model. The process is repeated $k$ times, each time a different subset is used for testing purposes.

\begin{figure*}
    \centering
    \includegraphics[width=0.97\textwidth]{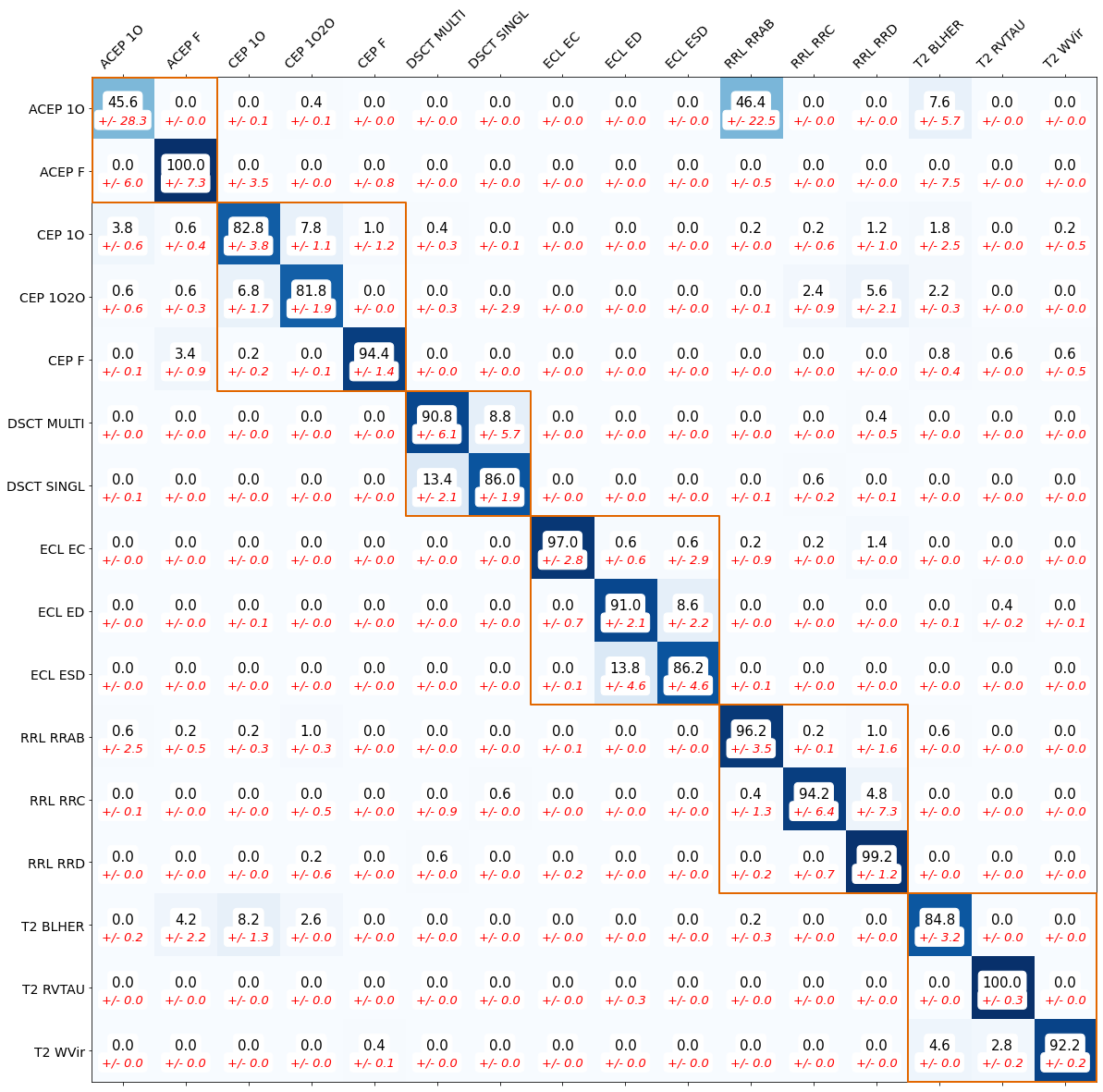}
    \caption{Confusion matrix of the test sample that summaries the prediction results of $16$ variable star sub-types. For this task, the Multiple-input Neural Network was used with light curve images and periods as inputs. The individual deviation values from the K-Fold cross-validation are shown with red color below the prediction values. For better readability the main variable star classes are marked with orange boxes.}
    \label{fig:subtypeConfmatrix}
\end{figure*}

To receive a statistically meaningful result, we performed a 10-fold cross-validation on the variable star data set with sub-type labels. The images and numerical parameters within each sub-type was split into 10 non-overlapping parts, each containing 500 light curves and periods. From these ten data packages, 4 500 inputs were used for training and 500 were used for testing purposes. After performing the cross-validation, we calculated the mean and standard deviation of the accuracies to characterize how well our neural network works. The results are listed in Table \ref{Tab:kfold} and visualized in Figure \ref{fig:subtypeConfmatrix}.

\section{Results and Discussion}
\label{sec:results_and_discussion}
In the following, we evaluate the training performance, present our classification results on the test sample of the main- and sub-types of variable stars, separately. Moreover, we examine the performance of the network for stars with known distances in the LMC.

\subsection{Training performance}
Figure\,\ref{fig:models_acc_loss_history} shows the evolution of accuracy and loss during the training and validation of the network as a function of epoch for both the CNN (left-hand side panel), and the Multiple-input Neural Network (right panel), and for the six main (top row) and sixteen sub-type variable star classes (bottom row). Running on the main groups, the Convolutional Neural Network achieved about 90 percent accuracy, and 140 epochs before the early stopping completed the teaching. If the same neural network was taught with 16 different sub-classes, although it was able to continue to run much longer, the accuracy did not change significantly from the 100th epoch, and in terms of end result it reached significantly lower accuracy than the previous one. The neural network produced the worst loss values in this run.
If additional input data was used, the curves smoothed out and the results improved. This is true for neural networks run on both main groups and subgroups. Beside some subtle anomalies, the accuracy increased and the loss decreased continuously. After 400 epochs, where the accuracy reached about 95\%, the loss started to flatten and shortly after this point the early stopping terminated the learning. As the training and validation metrics evolved in the same pace, the network did not overfit.
The figure also contains a comparison how well the neural network performs using the variable star main types or the the sub-classes. Although the performance is very similar, distinguishing between the 6 main type labels is more accurate than in the 16 sub-class case.

\subsection{Classification of the six main variable star classes}
\label{sec:multi_main}

Figure\,\ref{fig:multi_main} shows the confusion matrix of the classification results of the six main variable star types. Compared to the early test results from the plain Convolutional Neural Network without additional input parameters, like the period (see e.g. \citealt{Szklenaretal2020}), the first tests with additional numerical data showed greatly improved results regarding the identification of the six main variable star types. 

One can see in Figure \ref{fig:multi_main} that the accuracy of the well-represented  variable stars, e.g., RR Lyrae stars, is high ($\gtrsim~93\%$), while in case of ACeps, the classification result is $\sim 50\%$. The accuracy for most main types is around or over 90\%. About 9\% of the Type-II Cepheid test sample mix with classical Cepheids, a small amount (6\ RR Lyrae stars) mix with $\delta$ Scutis. Here false classification happened due to short period RR Lyrae stars and long period $\delta$ Scutis. 

The two most accurate classes are the eclipsing binaries and the $\delta$ Scutis, where almost every light curve were classified correctly. The classical Cepheid test light curves slightly mix with the RRLyr and Type-II Cepheids.

\begin{table}
\footnotesize
\centering
\caption{Aggregated results of the 10-Fold cross-validation in case of the training/validation and testing steps using our Multiple-input Neural Network. The individual deviation values can be seen in Figure~\ref{fig:subtypeConfmatrix}, embedded into the confusion matrix.  } 
\label{Tab:kfold}
\begin{tabular}{cccc}
\hline
& Mean accuracy & Std deviation & Best accuracy \\
\hline
\hline

Training    & 93.90\% & $\pm0.90\%$ & 95.38\% \\ 
Validation  & 93.23\% & $\pm0.78\%$ & 94.35\% \\ 

\hline
\end{tabular}
\end{table}

\subsection{Classification of sixteen variable star sub-classes}
\label{sec:multi_sub}

In order to test the efficacy and performance of our method we decided to apply our MINN to sub-classes, as well. Most of the variable star sub-types show explicitly distinguishable light curve shapes. The data set was split so that we could perform training for 16 different sub-types, which were selected from all the OGLE-III fields. Just like in the previous section, light curves and the period values were used as inputs. We used the same network architecture as for the main variable stars, but the number of predicted classes were changed to 16. The test results for this network are shown in Figure\,\ref{fig:subtypeConfmatrix}. We added the estimated scatter from the K-Fold cross-validation to the confusion matrix. 

Our main goal was to distinguish 16 different sub-types, we examined how these sub-classes perform within their main group. The different main variable star groups were marked with orange boxes in Fig.~\ref{fig:subtypeConfmatrix} and this shows that although there is scatter between the related sub-classes, but if we summarize the main types, the precision of the classification is over 90\%. 

Now we can see that the ACep 1O sub-type performs the worst and the test data's classification is mixing with the RRab group. This is due to the similar shape of the light curves and the periods.

The scatter of the accuracy for the ACep 1O variables is 22-28\%, meanwhile the scatter is lower (around 1-6\%) in case of the other classes.

There is a slight, 5.6\%  mix between the Cep 1O2O and the RRd subgroups. Both sub-types pulsate with two modes, but we phase folded every light curve only in the dominant mode, so these show significant distortion. 
An other significant mixing is between the short period BLHer sub-type and the classical Cepheid 1O group. 
The classification of the $\delta$ Scuti and ECL types is near perfect, if we look only at the whole main groups. But if we look at the sub-classes, distinguish the two $\delta$ Scuti sub-classes is difficult, the neural network confused about 10\% of the test data with the other subgroup. According to the eclipsing binaries, identifying the detached or semi-detached eclipsing binary stars is also a difficult task, the transition between these two groups is continuous \citep{Bodi_etal_binary}. 

\begin{figure*}
   \centering
   \includegraphics[width=\textwidth]{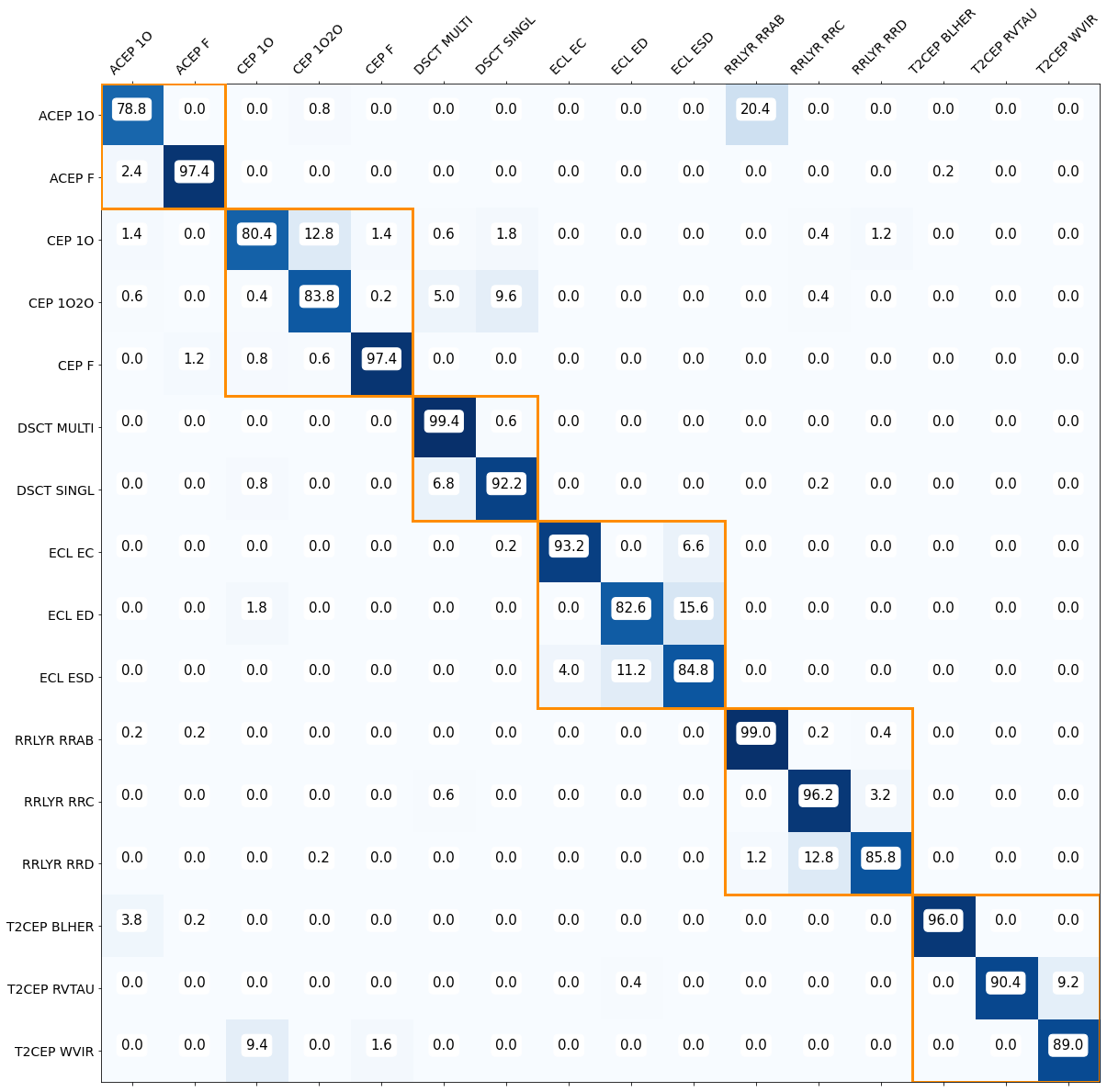}
     \caption{Confusion matrix of the testing phase, which shows the performance of our Multiple-input Neural Network using OGLE-III data only from the LMC field. The inputs were phase-folded light curves and the given star's period and Wesenheit-index. For better readability the main  variable star groups are marked with orange boxes. }
    \label{fig:lmc_training_result}
\end{figure*}

\subsection{Utilizing brightness for stars with known distances}
\label{sec:lmc}

As we demonstrated above, some subtypes are still hardly distinguishable based on their light curve shape and period, e.g. RRab and ACep stars. Here, we carry out an experiment, where additional parameters are fixed, and might be able to break this degeneracy. LMC comes to the rescue, where most of the variable (sub-)types are relatively well represented. Restricting ourselves to LMC only is beneficial, since it means that we can fix the distance, and the intrinsic luminosity difference will betray the various variable classes. 
Here we neglect the depth of the LMC, since the resulting error in luminosity introduced by this simplification is smaller than the intrinsic luminosity.

The OGLE-III database contains auxiliary information about each star, e.g. its magnitude, both in $I$ and $V$ bands, and the amplitude of variation, which can fed into our Multi-Input Neural Network as well. We calculated the reddening-free Wesenheit-index for each selected LMC variable star, and extended our neural network with this new input. The architecture of this network with an additional input can be seen in Figure\,\ref{fig:arch_imgPW}.

As the mixing of the ACep 1O variable star sub-class with the RRab stars is very high, we chose to discard every -- possibly -- foreground RR Lyrae stars that are brighter than 18 magnitude \citep[][]{OGLEIII_RRLYR_LMC}. Using this limit, 495 RR Lyrae stars were removed, this way we could reduce the degeneracy between RRab and ACep 1O stars, which have similar period.

\begin{figure}
   \centering
    \includegraphics[width=\columnwidth]{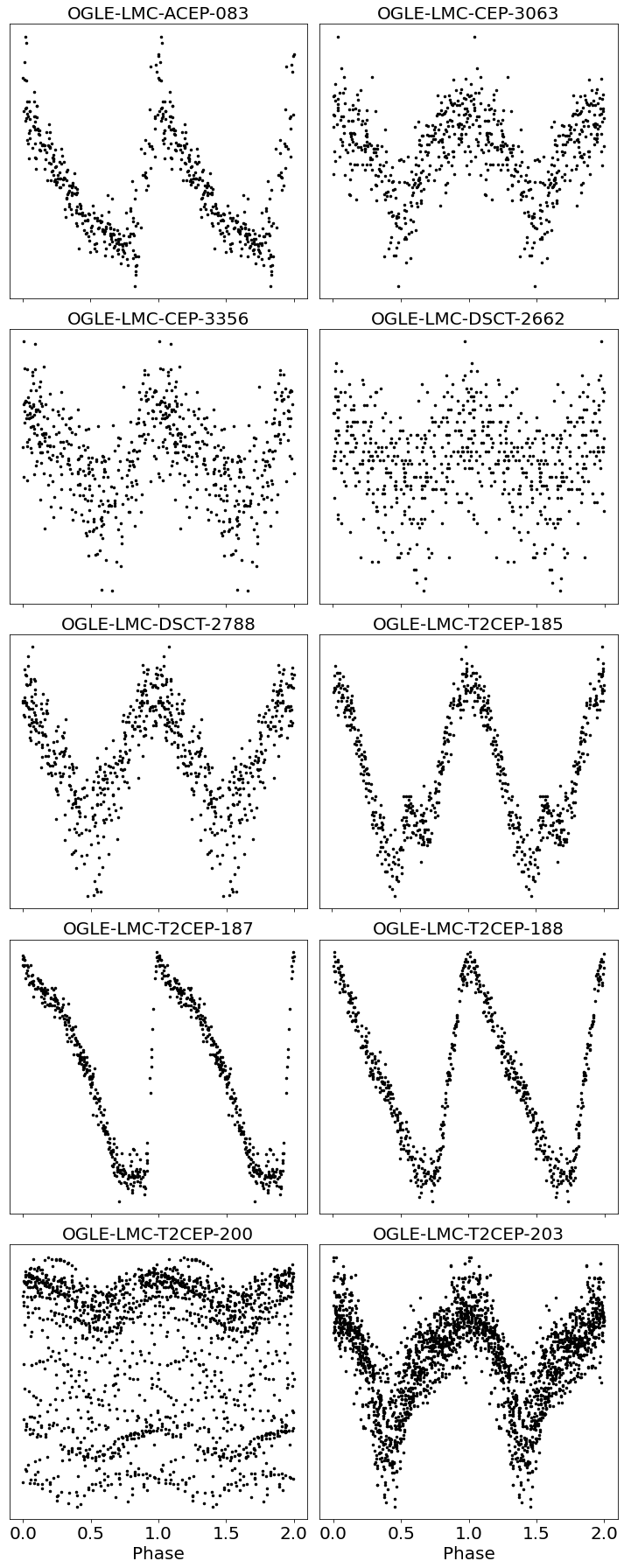}
    \caption{The light curves of the highlighted misclassified stars from Fig.~\ref{fig:PW_relation_full}. }
    \label{fig:wrong_class_lc}
\end{figure}

As there is less data in the LMC field only compared to the whole OGLE-III database, we had to generate additional artificial light curves for this training. The used variable star sub-types remained the same, 5\,000 light curves were used from each sub-type. Altogether we used 72\,000 light curves for the training and validation process and another 8\,000 light curves for testing purposes, without any overlapping between the two data sets. 

The confusion matrix on Figure~\ref{fig:lmc_training_result} contains our classification result of the test data from the OGLE-III LMC field. 
The classification result for each variable star was carefully checked, the misclassified stars were further investigated using the OGLE-III and OGLE-IV collection of variable stars. 

\begin{sidewaysfigure*}
\centering
    \includegraphics[width=\textwidth]{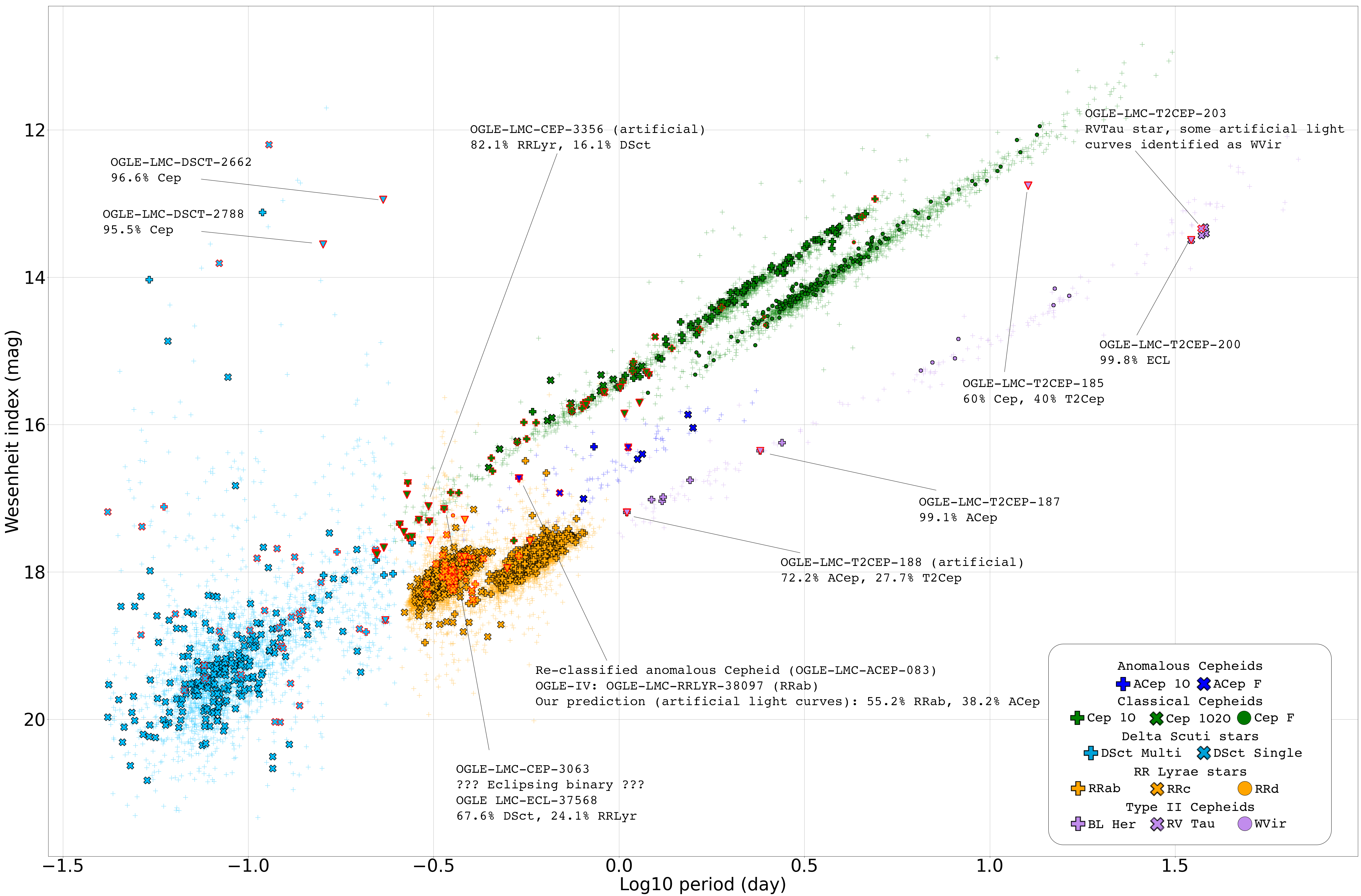}
    \caption{ Test result of the pulsating variable stars from the LMC field, based on their Period-Wesenheit-index relation. The training and the test was based on phase-folded light curves and the given star's period and Wesenheit-index. The neural network was trained on 16 variable star sub-types. For better readability we grouped them with one color for each main type and only the pulsating variable stars are shown in this figure. The crosses without border are from the training and validation, the correct classifications are shown with black borders. Those misclassified star which are plotted with red borders remained in their main group. The triangles with red edges are "true" misclassifications, which are found mostly on the borders of different pulsating variable star types.}
    \label{fig:PW_relation_full}
\end{sidewaysfigure*}

In Figure~\ref{fig:PW_relation_full} classification result of the 8000 test light curves is shown and to have better understanding about the test results, we illustrated these with three distinct groups: correctly classified stars, mixed stars within their own main group, and incorrectly identified stars. 
Figure~\ref{fig:wrong_class_lc} contains typical and interesting misclassified light curves, we will further investigate these stars below.

The removal of the foreground RRab stars greatly improved the neural networks accuracy by the distinction between of the ACep and RRLyr groups. The overall classification accuracy of the ACep variable star sub-classes improved greatly at least with 30\%. Still 20.4\% of the ACep 1O stars are classified as RRab by our MINN, but these misclassifications belong to the OGLE-LMC-ACEP-083, which was re-classified as an RRab star in the OGLE-IV database \citep[see][]{OGLEIV_ACEP}. Indeed, most of the artificial light curves generated from the ACEP-083 observations were classified as RRLyr stars.

Regarding the misclassifications of classical Cepheids, two groups can be mentioned. In the first group we can find stars with very short period, especially the 1O2O mode Cepheids, which mix with the $\delta$ Scutis. The stars in the other misclassified group are identified as ACep stars, these have a period about 1 day long and a characteristic, large amplitude light curve. 
According to the main group of classical Cepheids there is great mixing between the stars with different pulsation modes. About 13\% of the 1O mode Cepheids mix with the 1O2O pulsation mode stars.

The OGLE-LMC-CEP-3063 star was classified as a 1O classical Cepheid in the OGLE-III catalog, but has been reclassified as an eclipsing binary with a new ID in the OGLE-IV catalog \cite[OGLE-LMC-ECL-37568, see ][]{Soszynskietal2015}. Using the period, which is given in the OGLE-III catalog, the phase folded light curve indeed shows a continuous variation that resembles to a pulsating star. However, OGLE-IV lists a new period, which is twice the previous one. Using the latter, the phase curve immediately reveals alternating minima, which is characteristics of eclipsing binaries. As we used the OGLE-III data set, our classification has been misled by the wrong period, which is crucial in our case.

An other interesting case is the OGLE-LMC-CEP-3356, which has an uncertain classification in the OGLE-III and OGLE-IV catalogs as a possible RR Lyrae star \citep[Cep,][]{OGLEIII_CEP_LMC}. Our classification result shows 82.1\% probability that this star is an RRLyr.

The $\delta$ Scuti stars performed very well, just 1\% of the stars were not identified as $\delta$ Scuti. There is a significant mixing between the Singlemode and Multimode stars. 
We found two interesting cases. The first is OGLE-LMC-DSCT-2662, which is cataloged as an uncertain $\delta$ Scuti star in \cite{OGLEIII_DSCT}. Our classification shows 51.4\% probability for 1O Cepheid and 45.2\% for 1O2O Cepheid. It's apparent proximity to a bright star, very short period and a barely noticeable, 0.01 magnitude amplitude, therefore it is most probably a blend. OGLE-LMC-DSCT-2788 is much brighter (15.13mag in I-band) than the other data from the LMC field, which will probably makes it a galactic $\delta$ Scuti star. Our classification shows 64.9\% probability for a 1O Cepheid and 30.6\% for 1O2O Cepheid. 

The Wesenheit-index as additional input did not change much in the classificiation result of eclipsing binary stars. Although there is significant mixing between the sub-groups, but only 2\% of all test eclipsing stars were identified as a different variable star type.

Examining the results of the RRLyr stars we can conclude that almost all classifications belong to the main RRLyr type, thus mixing occurs mainly between the sub-groups only. The RRab and RRc sub-classes performed particularly well, but almost 13\% of the RRd sub-class is mixing between the RRc sub-group. This is understandable, since most of the RRd stars are overtone dominated, that is the amplitude of the radial overtone mode is higher than that of the fundamental mode. 

Regarding the misclassification of Type-II Cepheids the following cases can be mentioned. 
The W Vir star, OGLE-LMC-T2CEP-185 was classified as a Classical Cepheid with F pulsation mode (57.6\%). It is a relatively bright -- 14.5 magnitude (I-band) -- star, with  with a 12.7 day long period. This star was labeled as an outlier in the period-radius relation of Type-II Cepheids and was identified as a possible binary star by \cite{Groenewegen171,Groenewegen172}.

The BLHer type star, OGLE-LMC-T2CEP-187 was classified as an Anomalous Cepheid with 1O pulsation mode.
Another BLHer type star, OGLE-LMC-T2CEP-188 was classified as an Anomalous Cepheid by our MINN. The OGLE-IV catalog contains only 4 measurements in V-band. Because of this, the calculated V brightness shows 0.5 magnitude change compared to the OGLE-III data. If the Wesenheit-index would be calculated from this V-band measurement, the star would lie in the ACep region in the Period-Wesenheit relation.

There are three RV Tauri stars in the LMC, which show, beside pulsation, long-term mean brightness variation. To phase fold the light curves of these stars, we used the pulsation period, as this is given as the primary variability in the OGLE catalog.
In case of the RVTau type star,  OGLE-LMC-T2CEP-200 the variation of the mean amplitude is so large that it makes the pulsation pattern unrecognizable after phase-folding the light curve, leaving us with a confident classification as an eclipsing binary, which subgroup contains noisy light curves with very long periods.

OGLE-LMC-T2CEP-203 belongs to the RVTau sub-type, the original data was correctly classified as an RVTau star, yet misidentified most of the artificial light curves generated from this star as a WVir. Probably because the light curve's shape is not stable in time and the generation of artificial light curves removed this information.

\section{Trained weights and code availability}
\label{sec:weights}
We decided to publish the codes and weight files obtained by the neural network on the website of our institute, so that other research groups can use them as well. 
The files are available from the following link: (\url{https://konkoly.hu/KIK/data_en.html#ML}) 

\section{Conclusions}
\label{sec:conclusions}

In this paper we trained, validated and tested a Multi-Input Neural Network (MINN), which consists of an image classifier Convolutional Neural Network and simple dense layers which are used to handle additional numerical input data.
The light curves from which the input images were generated were downloaded from the OGLE-III database with the corresponding physical parameters (periods, brightness). To have as much data as possible, we collected light curves from the LMC, SMC, Galactic bulge, and Galactic disk fields. Because of the highly unbalanced number of stars in different classes  we generated artificial light curves to have equal amount of data in each variable star type. The augmented light curves were sampled from the posterior distributions of Gaussian Process regression of real observations.

For the classification we tested two kinds of setups: a 6-class input where the data sets of the main variable classes were merged, and a 16-class input where all the sub-classes were handled separately. The test showed that utilizing the periods beside the images of phase-folded light curves significantly improves the classification results. From the previous 77--99\% \citep{Szklenaretal2020} we were able to improve the accuracy to 89-99\%. Nonetheless, in case of underrepresented variable star sub-types, the classification results are significantly worse than for the other types, even using the additional numerical data. The low number of known Anomalous Cepheids (ACeps) prevents us to compile a diverse training sample for the network -- even with augmented training data. This explains the poor performance of the network in the case of first overtone ACep variables (see Figs. \ref{fig:multi_main} \& \ref{fig:subtypeConfmatrix}).

As an experiment, we restricted the training sample only to the LMC field, where the distance of the variable stars were fixed. The extended neural network received three different inputs: the phase-folded light curves, the periods and the reddening-free brightness (Wesenheit-index). The intrinsic luminosity difference helped to distinguish the first overtone ACep and the RRab variable stars with higher precision.

To be able to handle such an extensive amount of data as the OGLE catalog we used a high-performance GPU-accelerated computer. The generation of artificial light curve images took the most of time in the project due to the low scalability of Gaussian Process regression. The training and testing of the neural network took about 1.5 hours, after which we saved the weight file that could be used for further prediction and classification of thousands of light curves within just a couple of seconds.

As the various sky survey programs generate such vast amount of data each night, astronomers need to develop methods to be able to identify the different celestial objects in a reliable, accurate and efficient way. 
These methods can be used not only by current sky surveys, e.g. the Zwicky Transient Facility \citep[][ZTF]{Masci_2018} which has a continuously growing data set of $\sim$1 billion light curves, but also by such future projects, like the Vera C. Rubin Observatory Legacy Survey of Space and Time \citep[][LSST]{Ivezic_2019}. 
We  note that these surveys will have to accumulate enough data to build their own training samples in order to use our method. Depending on the strategy of the survey, some of these will reach that point pretty soon, e.g. quasi-continuous space-based photometric surveys, like TESS or PLATO, while others, like LSST will have to wait for months-years to have sufficient number of data points for a given object. Also, in order to use our method, we need to know whether a given object is (periodic) variable star or not, we need to know its period period and even its phase for more accurate classification. Such auxiliary information will not always be delivered by the official pipeline of a given survey, but for example in the case of LSST, in-kind contributions and brokers might deliver such data.

Training samples tailored to the characteristics of specific surveys and well designed neural networks can greatly accelerate data analysis with high reliability. 
We believe that the method we have developed and the ones based on it will be capable of this task in the future.

\begin{acknowledgements}

This project has been supported by the Lend\"ulet Program  of the Hungarian Academy of Sciences, project No.  LP2018-7/2021, the NKFI KH-130526 and NKFI K-131508 grants, the Hungarian OTKA Grant No. 119993, the 2019-2.1.11-TÉT-2019-00056, the MW-Gaia COST Action (CA 18104) grants and the KKP-137523 'SeismoLab' \'Elvonal grant of the Hungarian Research, Development and Innovation Office (NKFIH).
K.V. is supported by the Bolyai J\'anos Research Scholarship of the Hungarian Academy of Sciences. 

Authors acknowledge the financial support  of the Austrian--Hungarian  Action  Foundation (95\"ou3, 98\"ou5, 101\"ou13). 
On behalf of \textit{"Analysis of space-borne photometric data"} project we thank for the usage of ELHK Cloud (\url{https://science-cloud.hu}) that significantly helped us achieving the results published in this paper.
\end{acknowledgements}

\software{Python \citep{numpy},
Numpy \citep{numpy},
Pandas \citep{pandas},
Scikit-learn \citep{scikit},
Tensorflow \citep{Tensorflow},
Keras \citep{Keras} }

\clearpage


\onecolumngrid

\twocolumngrid

\bibliographystyle{aasjournal}
\bibliography{references}{}



\end{document}